\documentclass[%
	conference,
]{IEEEtran}

\def\COPYRIGHTYEAR{2018}
\def\CONFERENCE{2018 IEEE 87th IEEE Vehicular Technology Conference (VTC-Spring)} % set after the paper has been accepted
\def\DOI{10.1109/VTCSpring.2018.8417753}	% set after the paper has been published
\def\bibtex
{
	@InProceedings\{Sliwa/etal/2018b,\\
	author    = \{Benjamin Sliwa and Thomas Liebig and Robert Falkenberg and Johannes Pillmann and Christian Wietfeld\},\\
	title     = \{Efficient machine-type communication using multi-metric context-awareness for cars used as mobile sensors in upcoming {5G} networks\},\\
	booktitle = \{2018 IEEE 87th Vehicular Technology Conference (VTC-Spring)\},\\
	year      = \{2018\},\\
	address   = \{Porto, Portugal\},\\
	month     = \{Jun\},\\
	note      = \{\{Best Student Paper Award\}\},\\
	\}
}

\usepackage{graphicx}
\usepackage{epstopdf}
\usepackage{standalone}
\usepackage[nolist ]{acronym}

\usepackage{pdfpages}
\usepackage{tikz}
\usetikzlibrary{positioning}
\usepackage{hyperref}
\usepackage{pdfcomment}
\usepackage{hologo}

\usepackage{xstring}

\usepackage[utf8]{inputenc}
\usepackage{marvosym}

\usepackage{algorithm}
\usepackage{algpseudocode}
\usepackage{tikz}

\usepackage{pgfplots}
\pgfplotsset{compat=newest}
\pgfplotsset{
    box plot/.style={
        /pgfplots/.cd,
        fill=blue!30,
        only marks,
        mark=-,
        mark size=0.2em,
        /pgfplots/error bars/.cd,
        y dir=plus,
        y explicit,
    },
    box plot box/.style={
        /pgfplots/error bars/draw error bar/.code 2 args={%
            \draw  ##1 -- ++(.2em,0pt) |- ##2 -- ++(-.2em,0pt) |- ##1 -- cycle;
        },
        /pgfplots/table/.cd,
        y index=2,
        y error expr={\thisrowno{3}-\thisrowno{2}},
        /pgfplots/box plot
    },
    box plot top whisker/.style={
        /pgfplots/error bars/draw error bar/.code 2 args={%
            \pgfkeysgetvalue{/pgfplots/error bars/error mark}%
            {\pgfplotserrorbarsmark}%
            \pgfkeysgetvalue{/pgfplots/error bars/error mark options}%
            {\pgfplotserrorbarsmarkopts}%
            \path ##1 -- ##2;
        },
        /pgfplots/table/.cd,
        y index=4,
        y error expr={\thisrowno{2}-\thisrowno{4}},
        /pgfplots/box plot
    },
    box plot bottom whisker/.style={
        /pgfplots/error bars/draw error bar/.code 2 args={%
            \pgfkeysgetvalue{/pgfplots/error bars/error mark}%
            {\pgfplotserrorbarsmark}%
            \pgfkeysgetvalue{/pgfplots/error bars/error mark options}%
            {\pgfplotserrorbarsmarkopts}%
            \path ##1 -- ##2;
        },
        /pgfplots/table/.cd,
        y index=5,
        y error expr={\thisrowno{3}-\thisrowno{5}},
        /pgfplots/box plot
    },
    box plot median/.style={
        /pgfplots/box plot
    },
    boxplot/every median/.style={
    	ultra thick,dashed,cyan
    }
}

\definecolor{flexicolor}{RGB}{46,49,146}
\definecolor{amaricolor}{RGB}{237,28,36}

\usepackage{xspace}

\usepackage[binary-units=true]{siunitx}
\usepackage{psfrag}
\usepackage{graphicx}
\usepackage{tabularx,booktabs}
\usepackage{multirow}
\usepackage{rotating}

\renewcommand{\baselinestretch}{1}

\usepackage{multirow}

\usepackage[cmex10]{amsmath}
\usepackage{array}
\usepackage[caption=false,font=footnotesize]{subfig}
\usepackage{stfloats}
\renewcommand{\baselinestretch}{0.92}

\begin{document}

\newcommand{\paperTitle}{Efficient Machine-type Communication using Multi-metric Context-awareness for Cars used as Mobile Sensors in Upcoming 5G Networks}
\newcommand{\paperAuthors}{Benjamin Sliwa and Christian Wietfeld}
\newcommand{\paperEmails}{$\{$Benjamin.Sliwa, Christian.Wietfeld$\}$@tu-dortmund.de}

\newcommand{\githubUrl}{\footnote{Available at https://github.com/BenSliwa/MTCApp} }
\newcommand{\figurePadding}{0pt}
\newcommand{\figureTopPadding}{\figurePadding}
\newcommand{\figureBottomPadding}{\figurePadding}

\newcommand{\dummy}[3]
{
	\begin{figure}[b!]  
		\begin{tikzpicture}
		\node[draw,minimum height=6cm,minimum width=\columnwidth]{\LARGE #1};
		\end{tikzpicture}
		\caption{#2}
		\label{#3}
	\end{figure}
}

\newcommand{\fig}[4]
{
	\begin{figure}[#1]  	
		\vspace{\figureTopPadding}
		\centering		  
		\includegraphics[width=1\columnwidth]{#2}
		\caption{#3}
		\label{#4}
		\vspace{\figureBottomPadding}	

	\end{figure}
}

\newcommand{\subfig}[3]
{
	\subfloat[#3]
	{
		\includegraphics[width=#2\textwidth]{#1}
	}
	\hfill
}

%\begin{acronym}

\acrodef{ANN}{Artificial Neural Network}
\acrodef{M5T}{M5 Decision Tree}
\acrodef{SVM}{Support Vector Machine}
\acrodef{LR}{Linear Regression}

	\acrodef{DMR}{Deadline Miss Ratio}
	\acrodef{H2H}{Human-to-human}	
	\acrodef{CAT}{Channel-aware Transmission}
	\acrodef{pCAT}{predictive CAT}
	
	\acrodef{RSRP}{Reference Signal Received Power}
		\def\RSRP{\ac{RSRP}\xspace}
	\acrodef{RSRQ}{Reference Signal Received Quality}
		\def\RSRQ{\ac{RSRQ}\xspace}
	\acrodef{CQI}{Channel Quality Indicator}
		\def\CQI{\ac{CQI}\xspace}
	\acrodef{SNR}{Signal-to-Noise-Ratio}
		\def\SNR{\ac{SNR}\xspace}
	\acrodef{RSSI}{Received Signal-Strength Indicator}
		\def\RSSI{\ac{RSSI}\xspace}

	\acrodef{HTTP}{Hypertext Transfer Protocol}

		\acrodef{CAT}{Channel-aware Transmission}
		\acrodef{pCAT}{predictive CAT}
		\acrodef{MTC}{Machine-Type Communication}
			\def\MTC{\ac{MTC}\xspace}
		
		\acrodef{V2V}{Vehicle-to-Vehicle}
		\acrodef{V2I}{Vehicle-to-Infrastructure}

		% Communication Technologies
		\acrodef{LTE}{Long Term Evolution}
			\def\LTE{\ac{LTE}\xspace}
		\acrodef{LTE D2D}{LTE Device-to-Device}
		\acrodef{UE}{User Equipment}
			\def\UE{\ac{UE}\xspace}
		\acrodef{eNB}{Evolved Node B}
		\acrodef{IoT}{Internet of Things}
			\def\IoT{\ac{IoT}\xspace}
		\acrodef{SDR}{Software-Defined Radio}
			\def\SDR{\ac{SDR}\xspace}
			\def\SDRs{\acp{SDR}\xspace}

		% Mesh Routing
		\acrodef{AODV}{Ad hoc On-demand Distance Vector}
		\acrodef{B.A.T.M.A.N.}{Better Approach To Mobile Adhoc Networking}
		\acrodef{OLSR}{Optimized Link State Routing}

		\acrodef{OFDM}{Orthogonal Frequency-Division Multiple Access}

		% Evalutation
		\acrodef{KPI}{Key Performance Indicator}
		\acrodef{PDR}{Packet Delivery Ratio}

		% Network types
		\acrodef{MANET}{Mobile Ad-hoc Network}
		\acrodef{VANET}{Vehicular Ad-hoc Network}
		\acrodef{WMN}{Wireless Mesh Network}
		\acrodef{WSN}{Wireless Sensor Network}

		% Routing
		\acrodef{RREP}{Route Reply}
		\acrodef{RREQ}{Route Request}

		% Simulation Tools
		\acrodef{OMNeT++}{Objective Modular Network Testbed in C++}
		\acrodef{SUMO}{Simulation of Urban Mobility}
		\acrodef{Veins}{Vehicles in Network Simulation}
		\acrodef{NS2}{Network Simulator 2}
		\acrodef{NS3}{Network Simulator 3}
		\acrodef{LIMoSim}{Lightweight ICT-centric Mobility Simulation}

		% Network Layer
		\acrodef{IPv4}{Internet Protocol version 4}

		\acrodef{QoS}{Quality of Service}
		\acrodef{TOS}{Type of Service}
		
		% Link Layer
		\acrodef{MAC}{Medium Access Control}

		% Transport Layer
		\acrodef{UDP}{User Datagram Protocol}
		\acrodef{TCP}{Transmission Control Protocol}
		\acrodef{MTU}{Maximum Transmission Layer}

		% Vehicle classes
		\acrodef{UAV}{Unmanned Aerial Vehicle}
		\acrodef{UGV}{Unmanned Ground Vehicle}

%\end{acronym}

\acresetall
\title{\paperTitle}

\newcommand\Mark[1]{\textsuperscript#1}

\author{
	
	\IEEEauthorblockN{
		\textbf{Benjamin Sliwa\Mark{1}, Thomas Liebig\Mark{2}, Robert Falkenberg\Mark{1}, Johannes Pillmann\Mark{1} and Christian Wietfeld\Mark{1}}
	}

	\IEEEauthorblockA{
		\Mark{1}Communication Networks Institute, 
		\Mark{2}Department of Computer Science VIII\\ 
		TU Dortmund University, 44227 Dortmund, Germany\\
		e-mail:  $\{$Benjamin.Sliwa, Thomas.Liebig, Robert.Falkenberg, Johannes.Pillmann, Christian.Wietfeld$\}$@tu-dortmund.de 			
	}
}

\maketitle
%
% Make your adjustments here
%
\def\COPYRIGHTYEAR{2018}
\def\CONFERENCE{2018 IEEE 87th IEEE Vehicular Technology Conference (VTC-Spring)} % set after the paper has been accepted
\def\DOI{10.1109/VTCSpring.2018.8417753}	% set after the paper has been published

\def\bibtex
{
	@InProceedings\{Sliwa/etal/2018b,\\
	author    = \{Benjamin Sliwa and Thomas Liebig and Robert Falkenberg and Johannes Pillmann and Christian Wietfeld\},\\
	title     = \{Efficient machine-type communication using multi-metric context-awareness for cars used as mobile sensors in upcoming {5G} networks\},\\
	booktitle = \{2018 IEEE 87th Vehicular Technology Conference (VTC-Spring)\},\\
	year      = \{2018\},\\
	address   = \{Porto, Portugal\},\\
	month     = \{Jun\},\\
	note      = \{\{Best Student Paper Award\}\},\\
	\}
}
\ifx\CONFERENCE\VOID
\def\conferencenotice{Submitted for publication}
\def\copyrightnotice{}
\else
\ifx\DOI\VOID
\def\conferencenotice{Accepted for presentation in: \CONFERENCE}	
\else
\def\conferencenotice{Published in: \CONFERENCE\\DOI: \href{http://dx.doi.org/\DOI}{\DOI}
	
	\vspace{0.3cm}
	\pdfcomment[color=yellow,icon=Note]{\bibtex}    
	
}
\fi
\def\copyrightnotice{
	\copyright~\COPYRIGHTYEAR~IEEE. Personal use of this material is permitted. Permission from IEEE must be obtained for all other uses, including reprinting/republishing this material for advertising or promotional purposes, collecting new collected works for resale or redistribution to servers or lists, or reuse of any copyrighted component of this work in other works.
}
\fi
\def\overlayimage{%
	\begin{tikzpicture}[remember picture, overlay]
	\node[below=5mm of current page.north, text width=20cm,font=\sffamily\footnotesize,align=center] {\conferencenotice};
	\node[above=5mm of current page.south, text width=15cm,font=\sffamily\footnotesize] {\copyrightnotice};
	\end{tikzpicture}%
}
\overlayimage
\begin{abstract}
	
%
% Introduction
%
Upcoming 5G-based communication networks will be confronted with huge increases in the amount of transmitted sensor data related to massive deployments of static and mobile \ac{IoT} systems. Cars acting as mobile sensors will become important data sources for cloud-based applications like predictive maintenance and dynamic traffic forecast.
%
% Problem statement
%
Due to the limitation of available communication resources, it is expected that the grows in \ac{MTC} will cause severe interference with \ac{H2H} communication. Consequently, more efficient transmission methods are highly required. 
%
% Solution appraoch
%
In this paper, we present a probabilistic scheme for efficient transmission of vehicular sensor data which leverages favorable channel conditions and avoids transmissions when they are expected to be highly resource-consuming.
%
% Results
%
Multiple variants of the proposed scheme are evaluated in comprehensive real-world experiments. Through machine learning based combination of multiple context metrics, the proposed scheme is able to achieve up to 164\% higher average data rate values for sensor applications with soft deadline requirements compared to regular periodic transmission.

\end{abstract}

\IEEEpeerreviewmaketitle
\section{Introduction}

%
% Introduction
%
The deployment of autonomous cars will act as a catalyst for data-driven sensor systems. Crowdsensing-based applications such as road-roughness detection and distributed weather-forecasting will achieve significant gains in the availability and up-to-dateness of sensor data with the help of cars acting as mobile sensor nodes. Those kinds of applications do not have real-time requirements for the data, but \emph{soft deadlines}, where too old data is discarded.
%
% Problem statement
%
Consequently, upcoming 5G networks will require mechanisms for dealing with a high increase in the amount of transmitted \ac{MTC}-data. In addition to reliable data transfer, highly efficient data transmissions mechanisms are required to keep the interference with \ac{H2H} communication as low as possible.
%
% Solution approach vs state of the art
%
In earlier work, we have proposed the opportunistic transmission scheme \ac{CAT} \cite{IdeDuszaWietfeld2015} for \ac{LTE} cellular networks, which has been extended with a predictive component to consider future channel conditions with \ac{pCAT} in \cite{WietfeldIdeDusz2014}. Its general idea is to store sensor data locally as long as the channel quality is low in order to avoid packet loss and retransmissions. The buffered data is sent to a cloud-server when the device is experiencing a \emph{connectivity hotspot} with very high channel quality. The actual transmissions are triggered depending on a passive downlink channel quality indicator. In this paper, we generalize the principles and extend \ac{CAT} with a generalized multi-metric model for context-aware \ac{MTC} in vehicular environments. 
%
% Structure of the paper
%
The paper is structured as follows. After presenting relevant state-of-the-art work in this area, we present the analytical model of the proposed probabilistic transmission scheme and discuss methods for combining multiple transmission metrics based on generic approaches and means of machine learning. Afterwards, the setup of the experimental evaluation study is introduced and finally the results are presented. In order to guarantee a high level of reproducibility, we provide the developed measurement software in an Open Source way as well as the raw data obtained from our measurements. 
%
% Fig. Scenario
%
%\fig{b!}{fig/eps/scenario}{Application scenario: Cars as mobile sensors}{fig:scenario}

% MOTIVATION
% coexistence with other cell participants
% energy efficiency (ref Ide/Dusza)

% !TeX spellcheck = en_US
\section{Related Work}

%
% Available Metrics
%

Due to a constantly increasing amount of mobile traffic, anticipatory and opportunistic data transmission approaches gain greater attraction, especially for \MTC.
Therefore, researchers have started to excessively analyze the performance of real-life mobile networks to derive metrics and methods for the optimization of versatile mobile applications or the network itself at different system levels~\cite{BuiCesanaHosseiniEtAl2017}.

While many works address the optimization of the infrastructure side, e.g., optimizing the radio-resource scheduler by forecasting the \UE traffic~\cite{Abedini2014, Balachandran2013} and mobility~\cite{Calabrese2010, Margolies2016},
numerous approaches focus on the \UE side as the main trigger of transmissions in a cellular network:
With knowledge of the UE trajectory, centralized approaches like~\cite{Nicholson2008} incrementally generate maps of the link-quality at corresponding locations.
Such measurements, in conjunction with online-learning algorithms, may be used to derive coverage maps of a cellular network~\cite{Kasparick2016} and finally schedule transmissions according to the current connectivity.
Other researches propose decentralized approaches with the focus on QoS and QoE for continuous or buffered transmissions, e.g., video streaming~\cite{Zahran2016, Wang2016}. Here, the throughput-measurements of recent transmissions feed an adaption logic in the client to request a dynamically-scaled quality of the content according to the current link capacity.

For short transmissions, e.g. \MTC, passive metrics become more important since active probing is impractical in these cases~\cite{Halepovic2012}.
An empirical comparison of passive downlink indicators like \RSSI and \SNR on the throughput in different environments is addressed in~\cite{Wu2012}.
In~\cite{IdeFalkenbergKaulbarsEtAl2016} the authors analyze and identify the suitability of indicators like \RSRP and \RSRQ for estimations of the uplink connectivity at different cell loads.

\section{Solution Approach}

Regular transmission schemes for sensor data are mostly periodic and do not consider the current channel quality situation for timing the data transfer. Consequently, many transmissions are performed with a low transmission efficiency, require retransmissions and consume a lot of energy. Moreover, channel capacity is occupied for a long time period and is not available for other cell participants. It can be concluded, that the determination of favorable transmission times can have a significant impact on the overall system efficiency.
Consequently, \emph{connection hotspots} should be exploited whereas transmissions during \emph{connectivity valleys} should be avoided. In the following, we first present the analytical model of the proposed probabilistic transmission scheme. Afterwards different methods for combining the available indicators in order to achieve a better measurement for the actual channel quality are discussed.
%
% Fig. Example time behavior
%
\fig{}{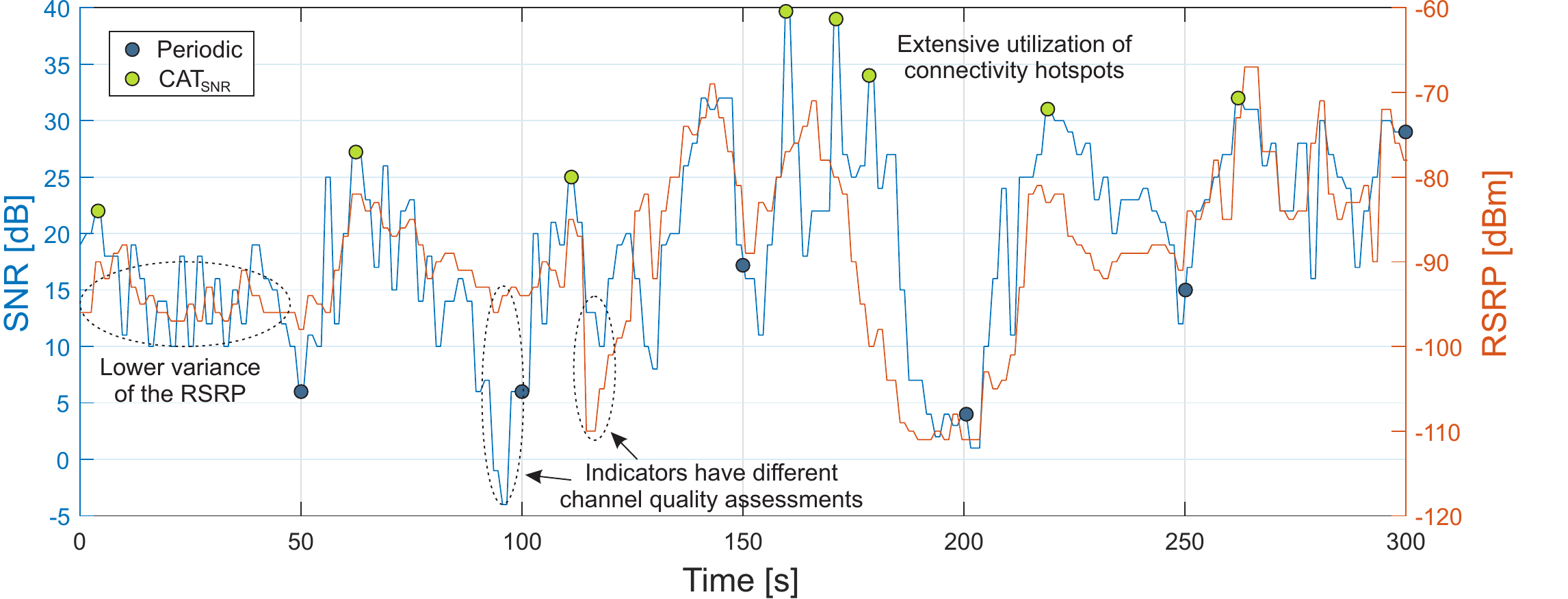}{Example temporal behavior for periodic data transfer and the proposed \ac{CAT}-based scheme. \ac{CAT} avoids transmissions during connectivity valleys and transmits more frequently when experiencing connectivity hotspots}{fig:time}
An example trace is shown in Fig.~\ref{fig:time}. While the periodic transmission approach sends data regardless of the current channel conditions, the proposed \ac{CAT} scheme avoids transmissions at bad channel conditions and sends data with a higher frequency if the channel quality is high.

\subsection{Analytical Model of the Transmission Scheme}

The generalized model for \ac{CAT} is shown in Eq.~\ref{eq:cat}. Each $t_{decision}$ seconds, it computes the transmission probability $p_{\Phi}(t)$ for an abstract metric $\Phi$ which is specified by a minimum value $\Phi_{min}$ and a maximum value $\Phi_{max}$. $\Phi$ is called \emph{conducive} if the channel quality gets better with higher values of $\Phi$ and \emph{harmful} if it gets worse. 
Transmissions are not performed if the time difference $\Delta t$ since the last transmission is below a minimum value $t_{min}$ and performed regardless of the current metric value if $\Delta t$ exceeds a defined maximum value $t_{max}$. The timeouts are used to specify the application requirements in terms of packet size and deadline requirements. If $\Delta t$ is in between those timeouts, a transmission probability is computed depending on the current metric value $\Phi(t)$ with the helper function $\Theta$ and Eq.~\ref{eq:sign}.
%
% Eq. 
%
\begin{align} \label{eq:cat} 
	p_{\Phi}(t) =\left\{\begin{array}{l} 
		0 : \Delta t \leq t_{min}\\	
		\Theta(\Phi(t))^{\alpha} : t_{min} < \Delta t < t_{max} \\ 
		1 :  \Delta t > t_{max} \\
	\end{array}\right.
\end{align}
\begin{align} \label{eq:sign} 
	\Theta \left( \Phi (t)\right)  =\left\{\begin{array}{ll} 
		\frac{\Phi(t) - \Phi_{min}}{\Phi_{max} - \Phi_{min}} & : \Phi \text{ is conducive} \\  
		1 - \left( \frac{\Phi_{max} - \Phi_{min}}{\Phi(t) - \Phi_{min}} \right)  & : \Phi \text{ is harmful}
	\end{array}\right.
\end{align}
The metric is further characterized using a weighting exponent $\alpha$ that controls the convergence speed of $\Phi$ towards $\Phi_{max}$ and specifies the impact of metric differences for the channel quality. Fig.~\ref{fig:analytic} shows the analytical transmission probability with respect to the current metric value for different exponent values.

%
% Fig: Analytical Transmission Probability
%
\fig{h}{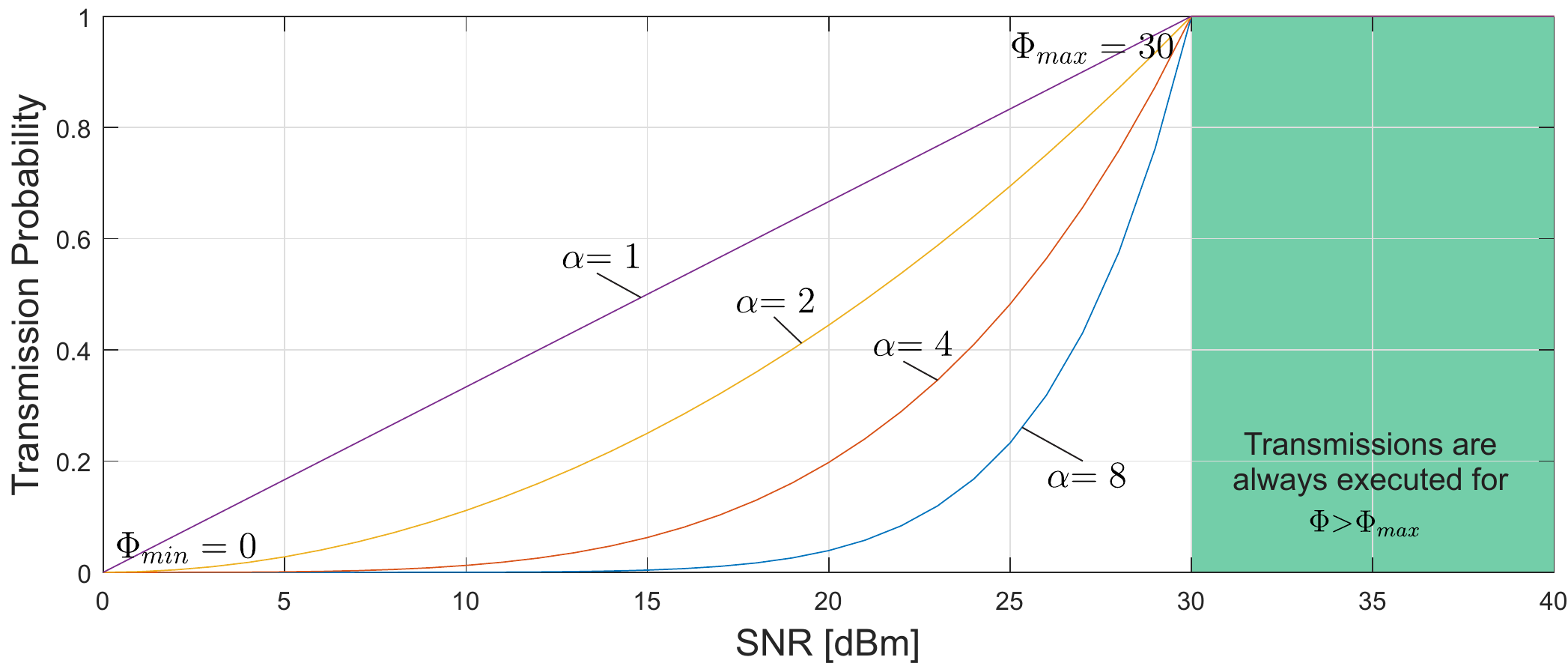}{Analytical transmission probability for \ac{CAT} for different values of the weighting exponent $\alpha$.}{fig:analytic}

\subsection{Multi-metric Transmission}

Cellular communication networks provide a wide range of different network quality indicators that can be used on the client side in order to evaluate the actual channel quality. In the following, different generic and machine-learning based approaches for combining the different indicators are described.

\subsubsection{Optimistic}

The transmission probability is computed for all considered metrics. As the resulting transmission probability, the maximum value is chosen (Eq.~\ref{eq:optimistic}) in order to leverage \emph{connectivity hotspots}.
%
% Eq. 
%
\begin{align} \label{eq:optimistic} 
	p_{opt}(t) &= max \left\lbrace p_{\Phi_{1}}(t), p_{\Phi_{2}}(t), ..., p_{\Phi_{n}}(t) \right\rbrace 
\end{align}

\subsubsection{Pessimistic}

The minimum value of all considered metric is used as the transmission probability. This approach focuses on avoiding transmissions during \emph{connectivity valleys}.
\begin{align} \label{eq:pessimistic} 
	p_{pes}(t) &= min \left\lbrace p_{\Phi_{1}}(t), p_{\Phi_{2}}(t), ..., p_{\Phi_{n}}(t) \right\rbrace 
\end{align}

\subsubsection{Weighted Mean}

If knowledge about the impact of the individual metrics on the overall performance is available, it can be used to strengthen or weaken their weight.
\begin{align} \label{eq:weighted_mean} 
	p_{mean}(t) &= \frac{1}{\sum g_{i}} \sum_{i=1}^{n} p_{\Phi_{i}}(t) \cdot g_{i}
\end{align}

\subsubsection{Machine Learning based Datarate Prediction}

With this approach, the anticipated data rate is computed and used as the decision metric for the transmission scheme. 
The processing pipeline for the machine learning process is illustrated in Fig.~\ref{fig:learning}. The available context parameters are used as the features and different learning models (\ac{ANN}, \ac{M5T}, \ac{SVM}, \ac{LR}) are applied (cf. Tab.~\ref{tab:ml-results} for the performance comparison). During the training phase, the measured effective data rate forms the label for the learning process. In contrast to the other multi-metric approaches, also the payload size of the current data packet is integrated into the transmission decision.
%
% Fig. Processing Pipeline for ML
%
\fig{b!}{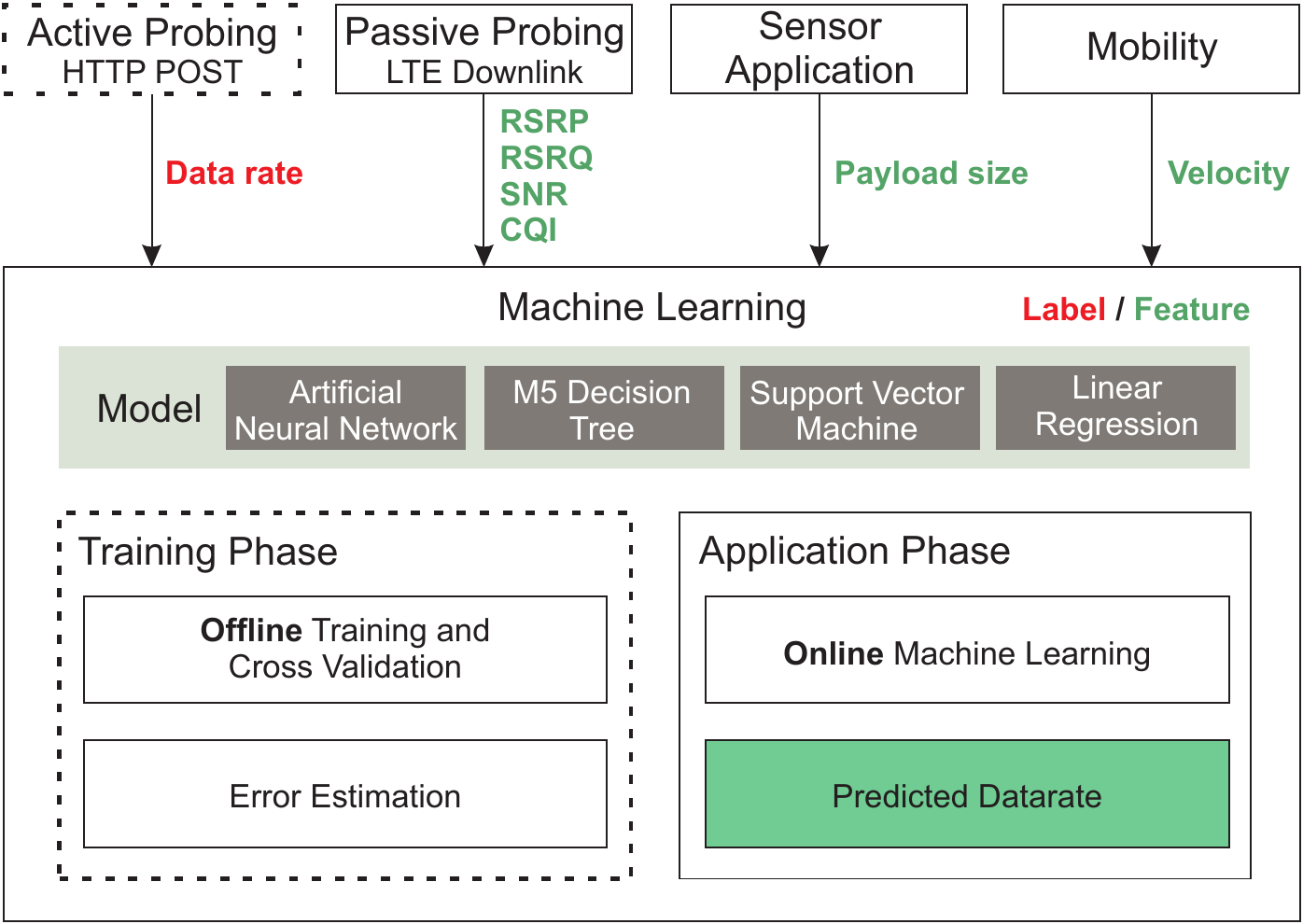}{Processing pipeline for the machine learning based datarate prediction process. The dashed components are only required for the training phase.}{fig:learning}
%
%
%
% vim: tw=78 ts=2 foldlevel=99 spell! wrap! lbr!:

% !TeX spellcheck = en_US
\section{Methodology}

In this section, we present the setup for our empirical study. The default parameters are denoted in Tab.~\ref{tab:simulation_parameters}. 
%
% Tab. Parameters
%
%\renewcommand{\arraystretch}{1.1}
\newcommand{\entry}[2]{#1 & #2 \\ \hline}
\newcommand{\head}[2]{\entry{\textbf{#1}}{\textbf{#2}}}

\begin{table}[h]

	\centering
	\caption{Parameters of the Reference Scenario}

	\setlength\extrarowheight{2pt}	% Some top-padding within cells, ugly otherwise
	\begin{tabularx}{\columnwidth}{X|X}
		\hline
		
		\head{Parameter}{Value}
		\entry{Sensor frequency}{1 Hz}
		\entry{Sensor payload size}{50 kByte}
		\entry{Transmission interval (periodic)}{30 s}
		\entry{$t_{decision}$}{1 s}
	
		\entry{$t_{min}$}{$\left\lbrace 10~s, 30~s\right\rbrace$}
		\entry{$t_{max}$}{120 s}

	\end{tabularx}
	\label{tab:simulation_parameters}

\end{table}
%
% Application
%
A virtual sensor application is generating data packets according to specified sensor frequency and packet size values. The packets are buffered until a transmission decision is made for the whole buffer and the data has been transmitted successfully. For the performance evaluation, a periodic scheme that uses $t_{min}$ is used as the transmission interval is used as a reference for the probabilistic model. 
%
% Fig. Map
%
\fig{}{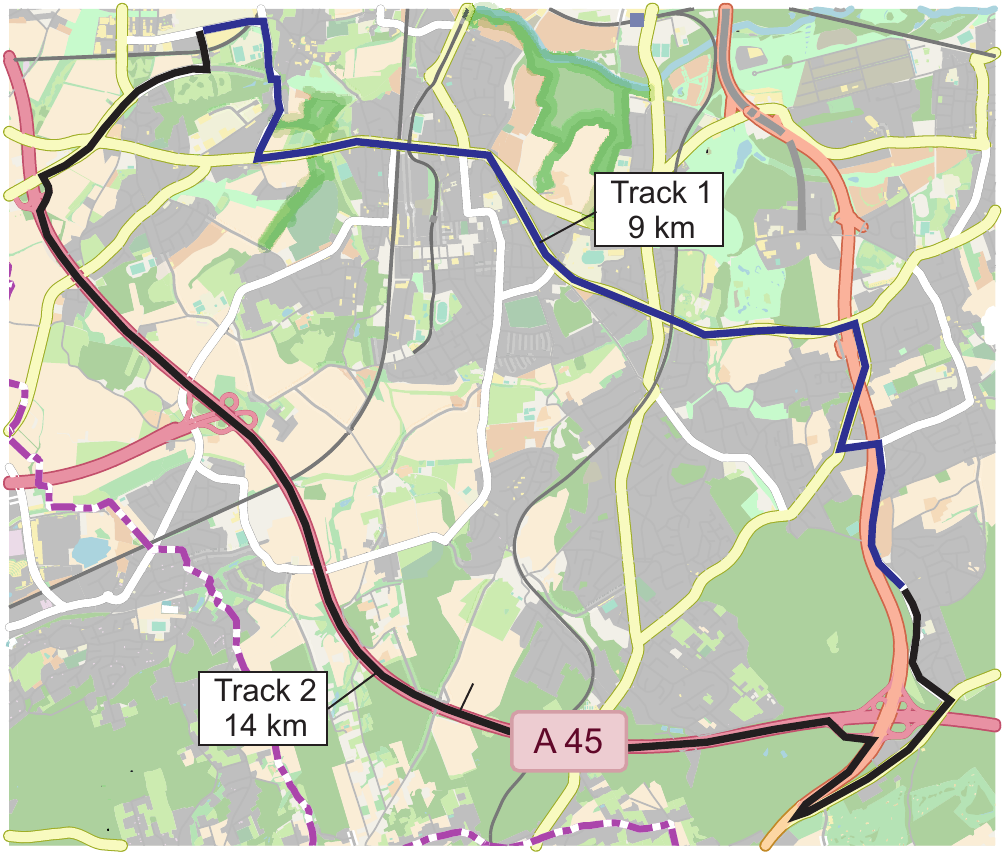}{Overview of the 23km long test route containing different street types and speed limitations. (Map: ©OpenStreetMap contributors, CC BY-SA)}{fig:map}
Fig.~\ref{fig:map} shows a map of the evaluation scenario and consists of two connected tracks with different characteristics. Track 1 is set up in a typical suburban area with speed limitations ranging from 30 km/h to 70 km/h. Track 2 is mostly highway with a speed limitation of 130 km/h. For most of the following evaluations, data is aggregated from both tracks and separate discussions of the tracks are performed when considered beneficial. For each parametrization of the transmission scheme, 5 runs were performed on each of the tracks. Overall, more than 4000 transmissions were performed on a total distance of more than 1000 km.
%
% Data Collection
%
The measurements were performed in a public \ac{LTE} network. For the collection of network quality data as well as mobility information, we developed an Android-based measuring app featuring live visualization that is available as Open Source \githubUrl. The measurements were performed with a Samsung Galaxy S5 Neo (Model SM-G903F). The raw data of the measurements is provided at \cite{Sliwa2017} as Open Access data.
%
% Tab. Metrics
%
%\renewcommand{\arraystretch}{1.1}
\renewcommand{\entry}[6]{#1 & #2 & #3 & #4 & #5 & #6\\ \hline}
\renewcommand{\head}[6]{\entry{\textbf{#1}}{\textbf{#2}}{\textbf{#3}}{\textbf{#4}}{\textbf{#5}}{\textbf{#6}}}

\begin{table}[h]

	\centering
	\caption{Parameters for the considered metrics}

	\setlength\extrarowheight{2pt}	% Some top-padding within cells, ugly otherwise
	\begin{tabularx}{\columnwidth}{X|X|X|X|X|X}
		\hline

		\head{}{$\Phi_{RSRP}$}{$\Phi_{RSRQ}$}{$\Phi_{SNR}$}{$\Phi_{CQI}$}{$\Phi_{M5T}$}
		\entry{min}{-120 dBm}{-11 dB}{0 dB}{2}{0 MBit/s}
		\entry{max}{-80 dBm}{-4 dB}{30 dB}{16}{$\left\lbrace 15,18\right\rbrace$  MBit/s}
		\entry{$\alpha$}{8}{6}{8}{6}{8}
		\entry{conducive}{true}{true}{true}{true}{true}

	\end{tabularx}
	\label{tab:metrics}

\end{table}
Tab.~\ref{tab:metrics} contains the parametrization for the considered metrics. For $\Phi_{RSRQ}$ and $\Phi_{CQI}$, the exponent $\alpha$ is chosen lower as the value range is smaller.
\section{Results}

In this work, the evaluation is focused on the application layer throughput (referred to as the \emph{goodput}) as a measurement for the transmission efficiency and the age of the sensor data. As discussed in \cite{IdeDuszaWietfeld2015}, higher throughput values significantly reduce the interference with other cell user as well as the overall energy consumption of the mobile device. 

\subsection{Single Metric Transmission Schemes}\label{sec:single_metric}

%
% Fig. Throughput vs Metrics
%
\begin{figure*}[!tbp] 
	\centering
	
	\subfig{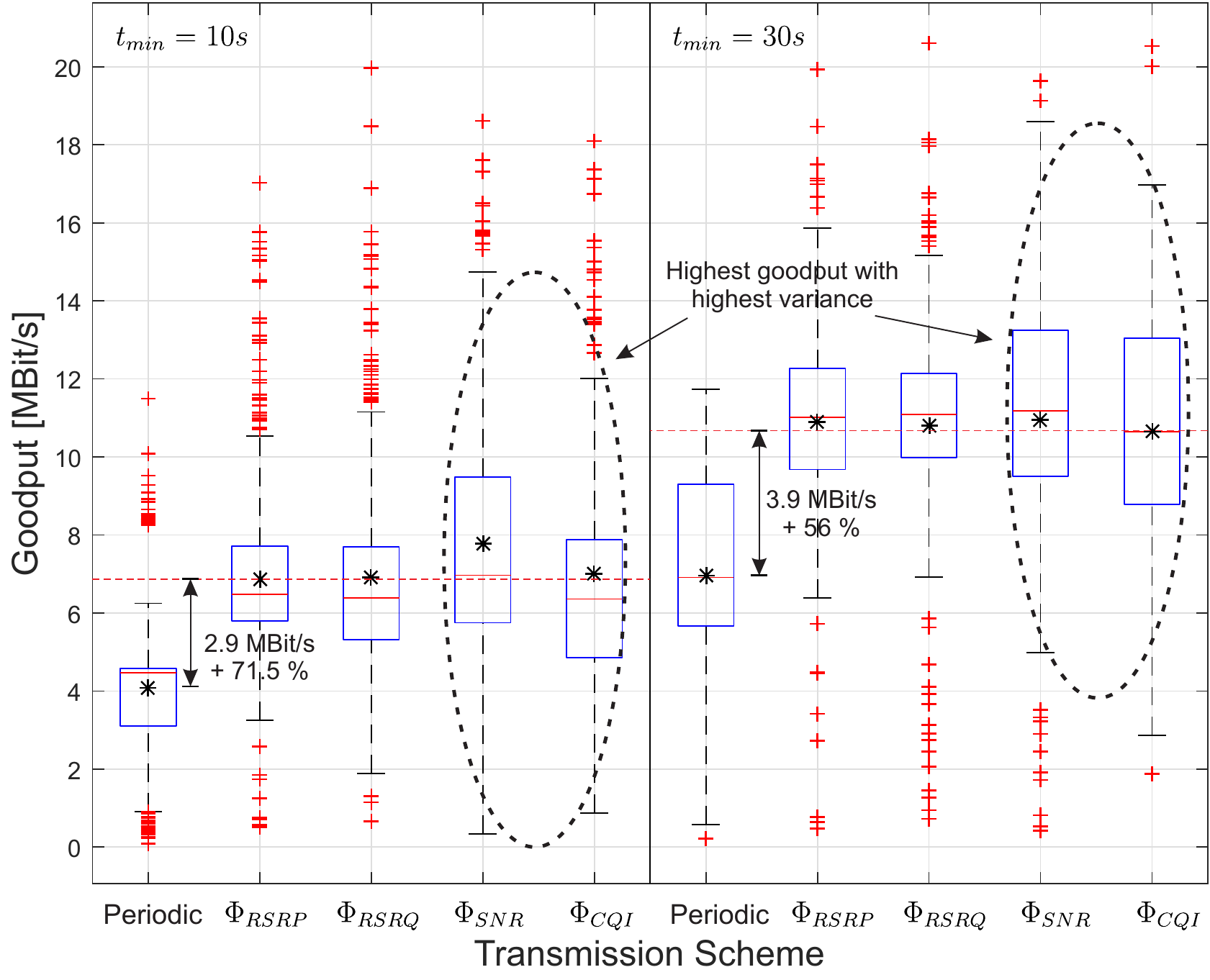}{0.45}{}
	\subfig{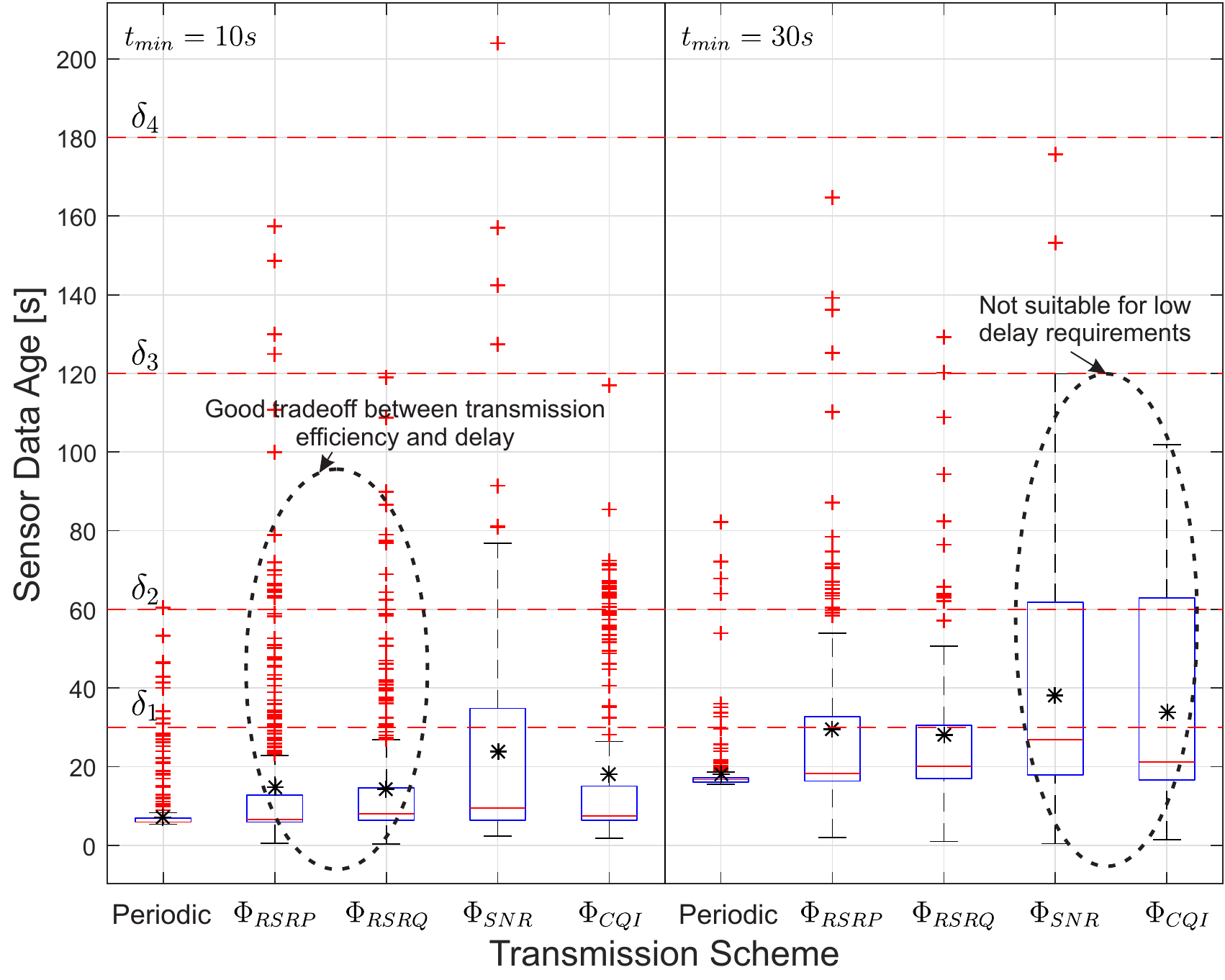}{0.45}{}
	
	\caption{End-to-end goodput and sensor data age for the considered base metrics}
	\label{fig:metrics}
	\vspace{-10pt}
\end{figure*}
The end-to-end goodput for the considered metrics is shown in Fig.~\ref{fig:metrics}. All \ac{CAT}-based metrics achieve significantly higher goodput values than the periodic scheme. While the highest gain is achieved for the \ac{SNR}-based metric, it also shows the highest variance. In many cases the reported \ac{SNR} changes significantly while the transmission is still being performed. 
%
% Tab. DMR
%
\begin{table}[h]
	\centering
	\caption{Resulting Deadline Miss Ratio for Multiple Application Requirements}
	\setlength\extrarowheight{2pt}	% Some top-padding within cells, ugly otherwise
	\begin{tabularx}{\columnwidth}{l|X|XXXXXXXX|X|}\hline
		
		Deadline & $t_{min}$ & Periodic & $\Phi_{RSRP}$ & $\Phi_{RSRQ}$ & $\Phi_{SNR}$ & $\Phi_{CQI}$ \\\hline
		
		\multirow{2}{*}{$\delta_{1}=30~s$} 
		& 10~s & 0.012 & 0.119 & 0.099 & 0.281 & 0.279 \\\cline{2-7}
		& 30~s & 0.034 & 0.285 & 0.263 & 0.464 & 0.48
		\\ \hline
		
		\multirow{2}{*}{$\delta_{2}=60~s$} 
		& 10~s & 0.001 & 0.046 & 0.029 & 0.139 & 0.186 \\\cline{2-7}
		& 30~s & 0.015 & 0.109 & 0.082 & 0.257 & 0.44
		\\ \hline
		
		\multirow{2}{*}{$\delta_{3}=120~s$}  
		& 10~s & 0 & 0.01 & 0 & 0.017 & 0 \\\cline{2-7}
		& 30~s & 0 & 0.021 & 0.012 & 0.014 & 0
		\\ \hline	
		
		\multirow{2}{*}{$\delta_{4}=180~s$} 
		& 10~s & 0 & 0 & 0 & 0.004 & 0 \\\cline{2-7}
		& 30~s & 0 & 0 & 0 & 0 & 0
		\\ \hline		
	\end{tabularx}
	\label{tab:dmr}
\end{table}
Since the selection of a transmission scheme is bound to the requirements of the actual application, the resulting \ac{DMR} values for three different generic applications are compared in Tab.~\ref{tab:dmr}. The deadlines were chosen with respect to realistic requirements of traffic management systems as discussed in \cite{VandenbergheVanhauwaertVerbruggeEtAl2012}.
If the application does not have strict deadline requirements, a higher $t_{min}$ value should be selected in order to increase the minimal packet size and achieve a better ratio of payload size and protocol overhead.
%
% Fig. Correlation
%
\begin{figure*}[] 
	\centering
	
	\includegraphics[width=.3\textwidth]{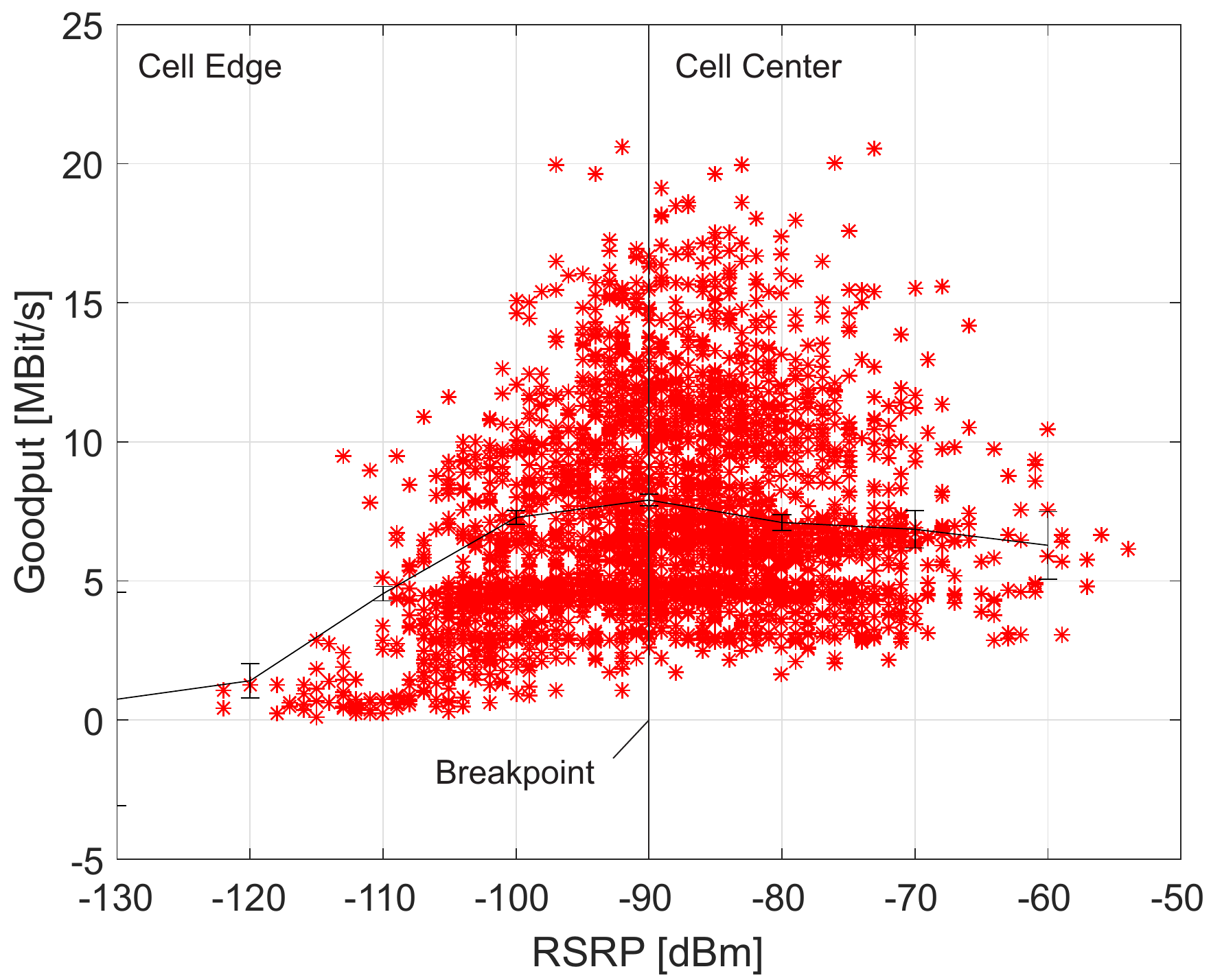}\quad
	\includegraphics[width=.3\textwidth]{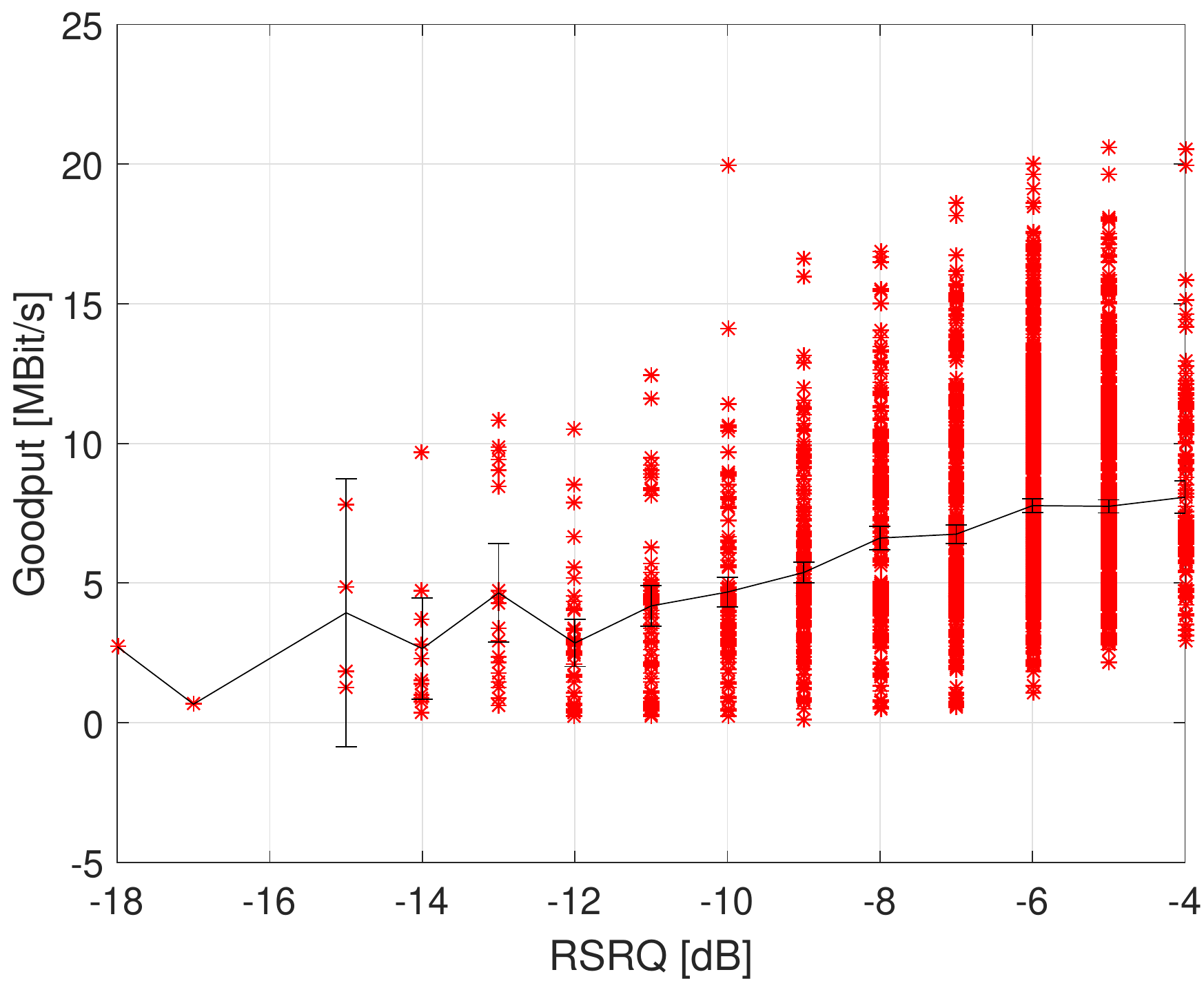}\quad
	\includegraphics[width=.3\textwidth]{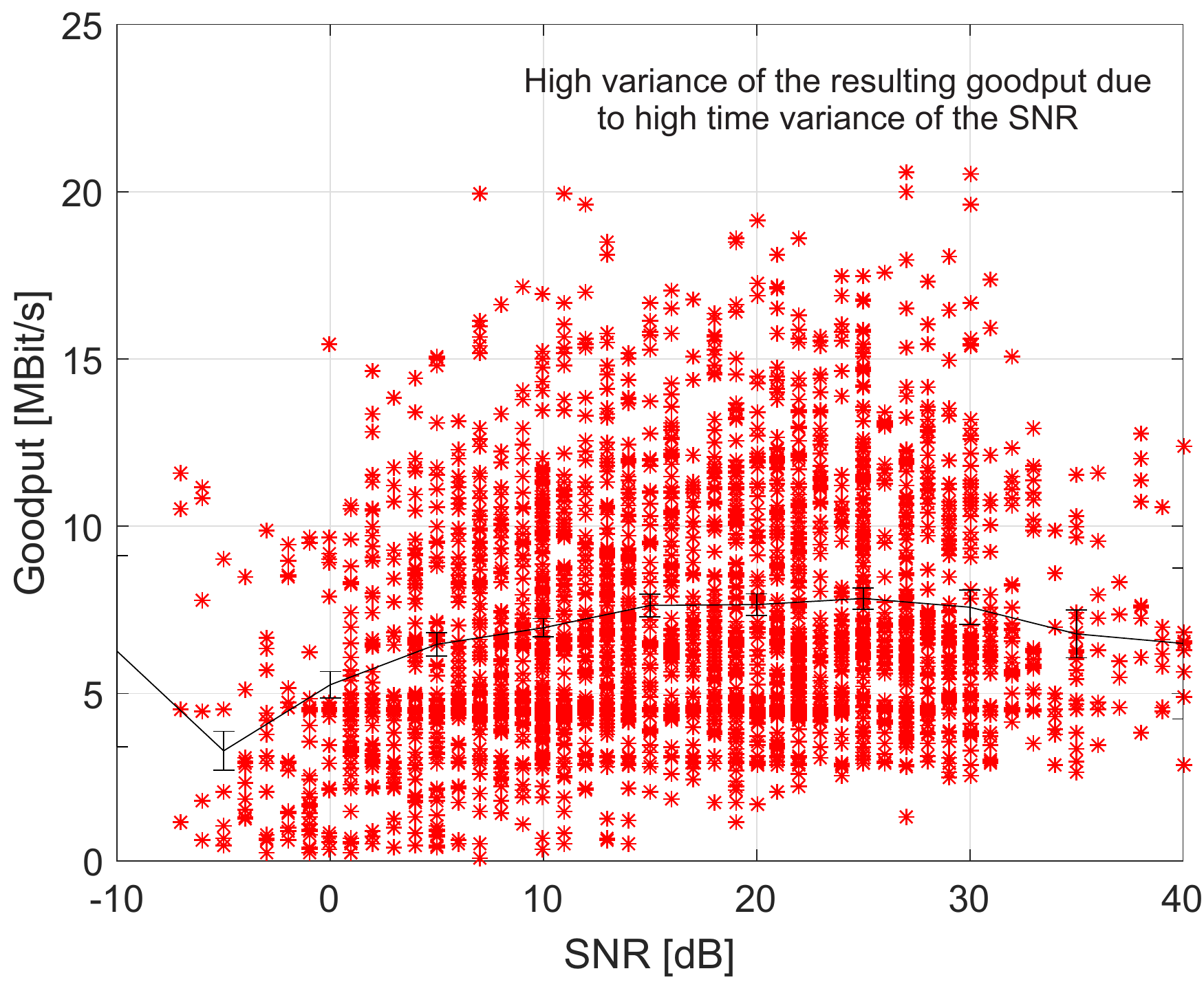}
	
	\medskip
	
	\includegraphics[width=.3\textwidth]{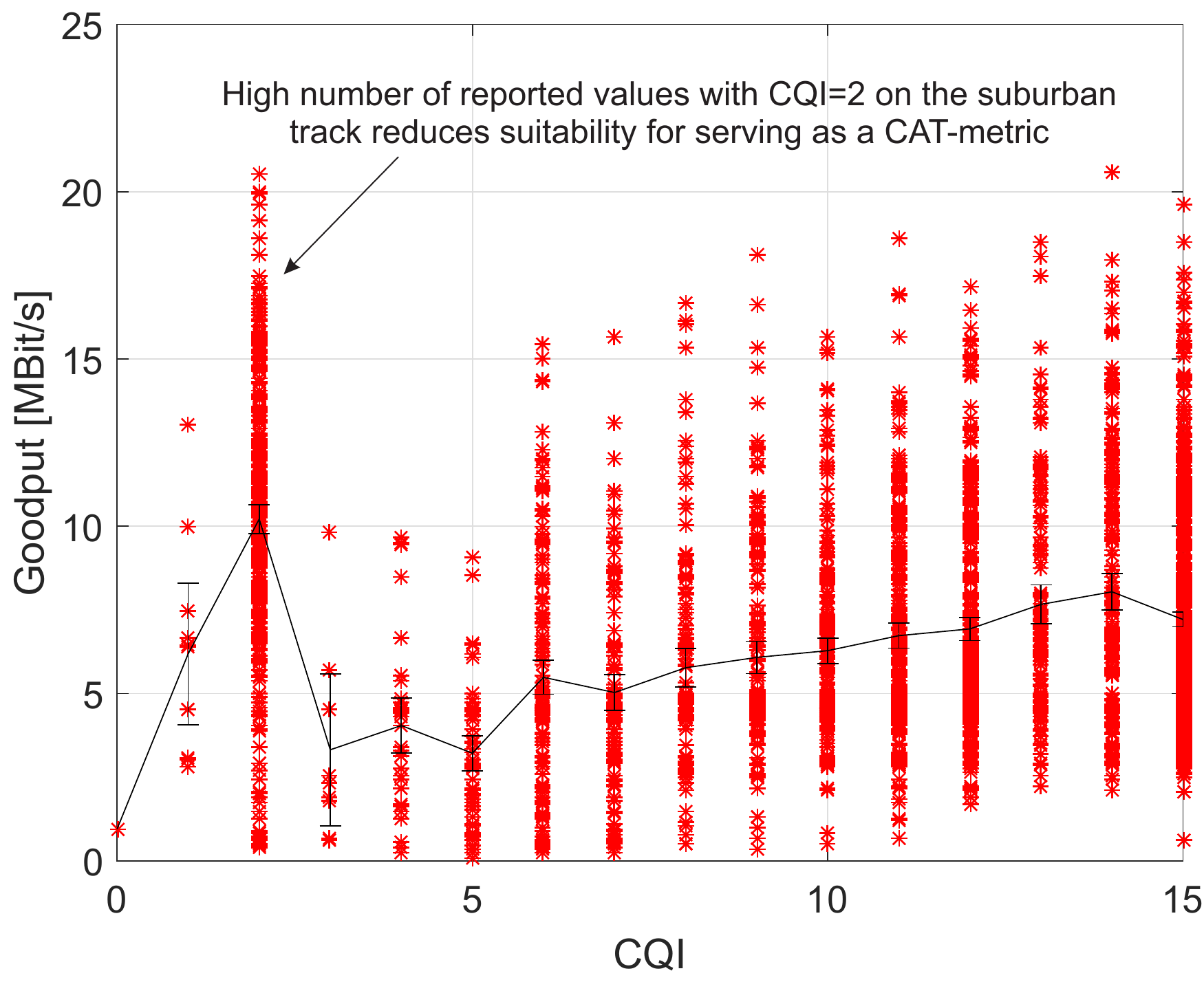}\quad
	\includegraphics[width=.3\textwidth]{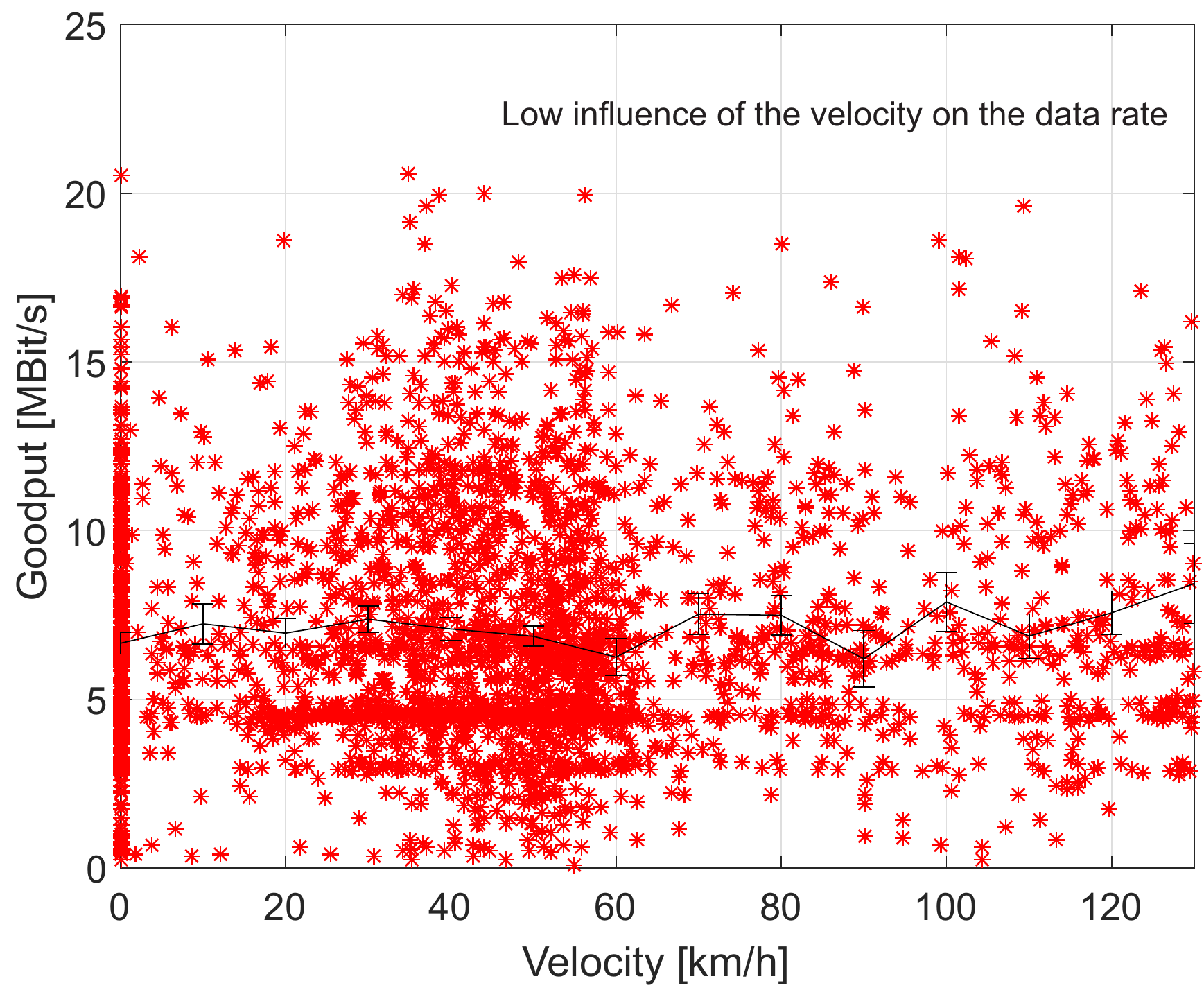}\quad
	\includegraphics[width=.3\textwidth]{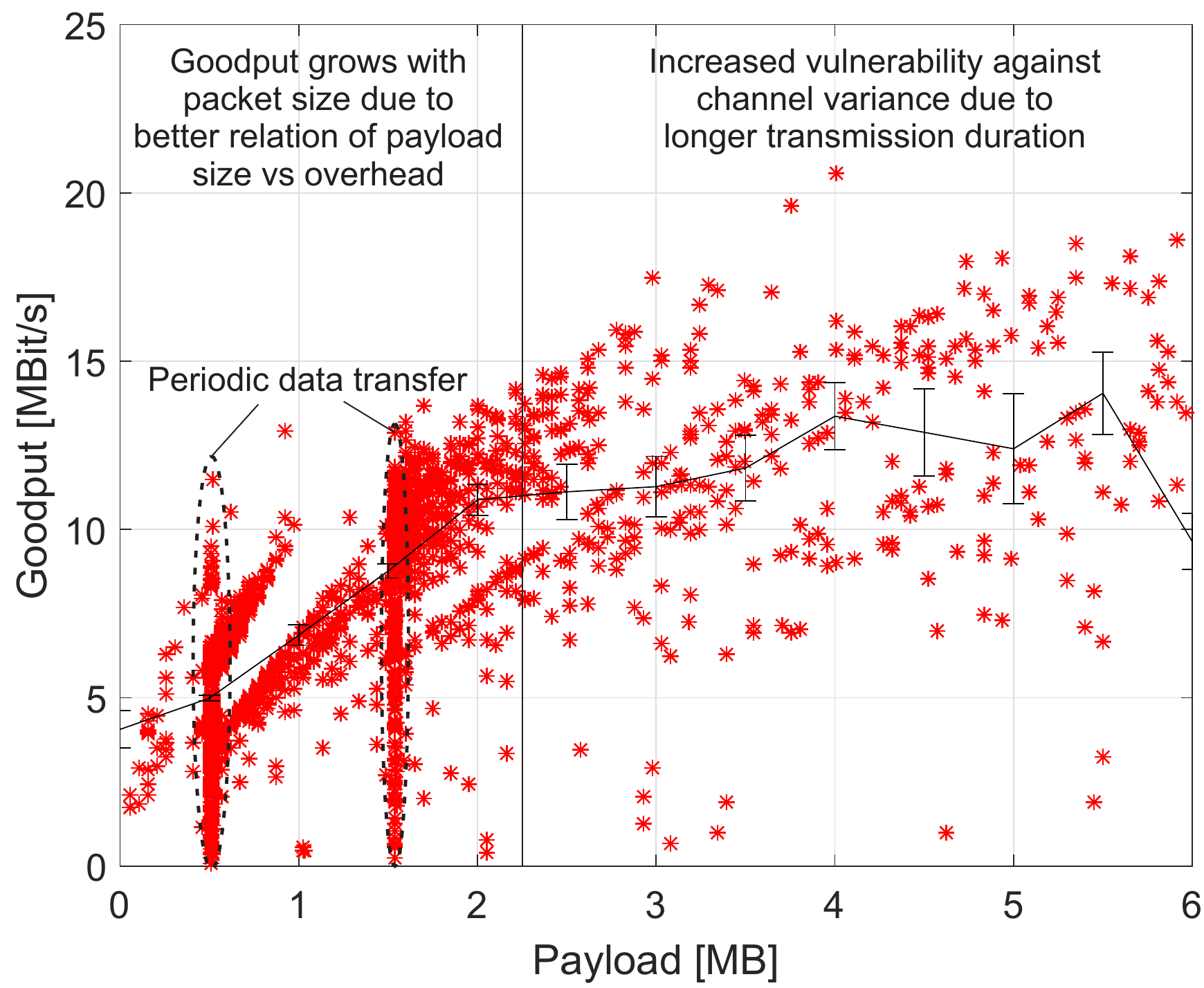}
	
	\caption{Correlation of different indicators with the resulting end-to-end goodput. The plots were generated from the data obtained from the single-metric \ac{CAT}-based transmissions as well as from the periodic transmissions. The black curve shows the 0.95 confidence interval of the mean value.}
	\label{fig:correlation}
	\vspace{-10pt}
\end{figure*}
The correlation between the considered metrics and the achieved goodput is shown in Fig.~\ref{fig:correlation} for the acquired data from Sec.~\ref{sec:single_metric}. 
%
% RSRP
%
The \ac{RSRP} is strongly correlated with the resulting data rate if the device is in the cell edge region. A breakpoint of the curve can be identified at -90 dBm that separates cell edge behavior from cell center behavior. After the breakpoint, its meaningfulness is reduced to the increased interference with other communicating devices.
%
% RSRQ
%
%
% SNR
%
Since the plots only consider the value of the \ac{KPI} for the beginning of the transmissions, its meaningfulness is reduced for longer transmission durations. This is even more dramatic if the \ac{KPI} itself has a high variance, as it is the case for the \ac{SNR}.
%
% CQI
%
%
% Velocity
%
%
% Payload Size
%
The packet size has a high impact on the resulting data rate as it controls the relation of payload and protocol overhead. Two  areas with different characteristics can be identified. Up to approximately 2.25 MB, the protocol overhead has a dominant influence and the goodput improves with higher packet sizes. For bigger packets, the variance of the data rate is highly increased, as the longer transmission duration increases dependency to the channel variance. In the plot, both periodic transmission schemes ($t_{min}=10, t_{min}=30$) can be clearly identified, as they do not alter the payload size of the packets.
As a result of the correlation considerations, no single metric is able to provide a robust measurement for the channel quality in all considered situations.

\subsection{Machine Learning Based Datarate Prediction}

Prediction of the datarate is a regression task. Thus, we train multiple regression methods and validate their performance in 10-fold cross validation. During the training phase, the transmission time and the payload size of the \ac{HTTP} POST transmissions is monitored and the passive quality indicators are recorded. The models we compare are Artificial Neural Networks, Support Vector Machine (using Polynomial Kernel), M5 decision tree and a simple linear regression. The performance of the models is measured using the test data fold in terms of correlation, mean absolute error and root mean squared error. Results can be found in Table~\ref{tab:ml-results}. The highest prediction accuracy is achieved for \ac{M5T} which is the only machine learning model which is considered for further evaluations. 
%
% M5T Performance
%
The performance of the prediction model is visualized in Fig.~\ref{fig:m5t-perform} where measured data rate and its predicted value are compared. For the proposed transmission scheme, only the overestimation of the channel quality is considered harmful as only a worst-case estimation is required and underestimation will most likely even increase the resulting data rate.
%
% Tab. Learning Performance
%
\begin{table}[h]
	
	\centering
	\caption{Performance of regression models: Artificial
		Neural Network (ANN), Support Vector Machine (SVM), M5 decision
		Tree (M5T), Linear Regression (LR) compared in terms of Correlation, mean
		absolute error (MAE) and root mean squared error (RMSE). Best values are
		highlighted.}
	
	\setlength\extrarowheight{2pt}	% Some top-padding within cells, ugly otherwise
	\begin{tabularx}{\columnwidth}{l|X|X|X}
		
		\hline \bf{Model} & \bf{Correlation} & \bf{MAE} & \bf{RMSE} \\ 
		\hline ANN & 0.83 & 1.62 & 2.07 \\ 
		\hline SVM & 0.80 & 1.61  & 2.13 \\ 
		\hline M5T & \bf{0.86} & \bf{1.33} & \bf{1.81}  \\
		\hline LR &  0.57 & 2.19 & 2.92 \\
		\hline 	
		
	\end{tabularx}
	\label{tab:ml-results}

\end{table}
%
% Fig. M5-Tree Performance
%
\fig{}{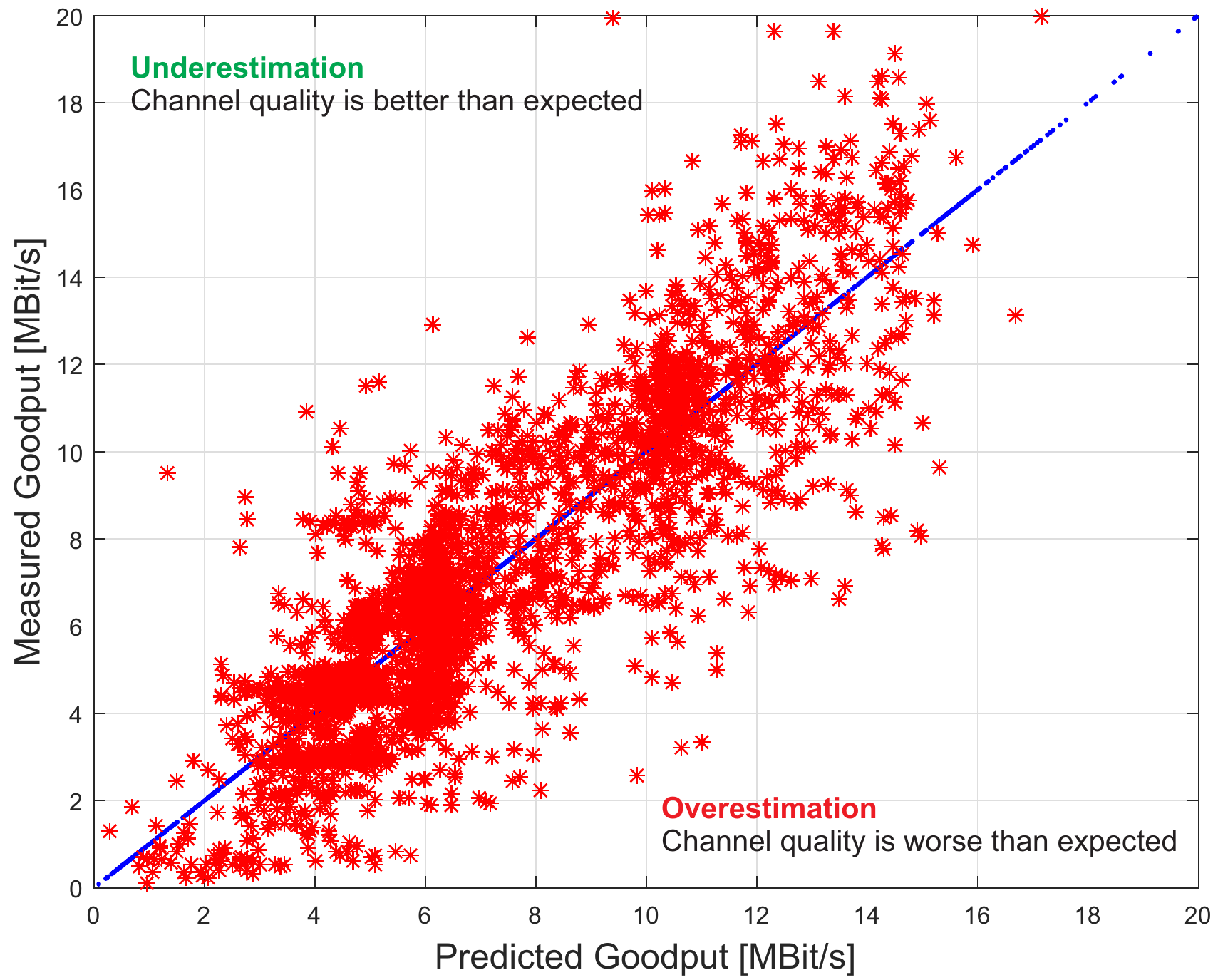}{Measured data rate versus predicted data rate using the M5 decision tree. Diagonal is highlighted in blue. Underestimations of the channel quality are not considered harmful for the proposed transmission scheme as it only requires a worst-case estimation.}{fig:m5t-perform}

\subsection{Multi-metric Transmission Schemes}

%
% Fig. Throughput vs Metric
%
\begin{figure*}[] 
	\centering
	
	\subfig{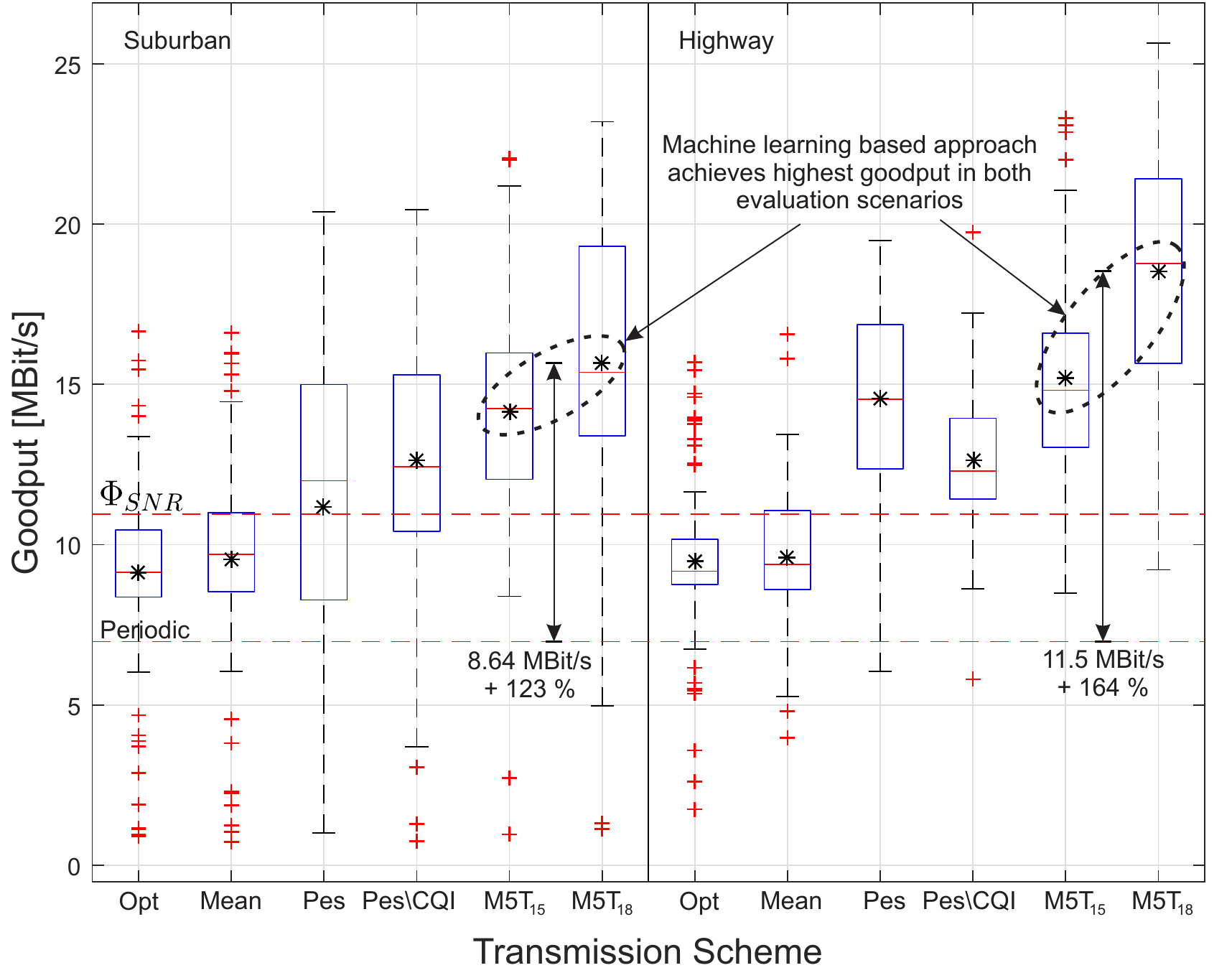}{0.45}{}
	\subfig{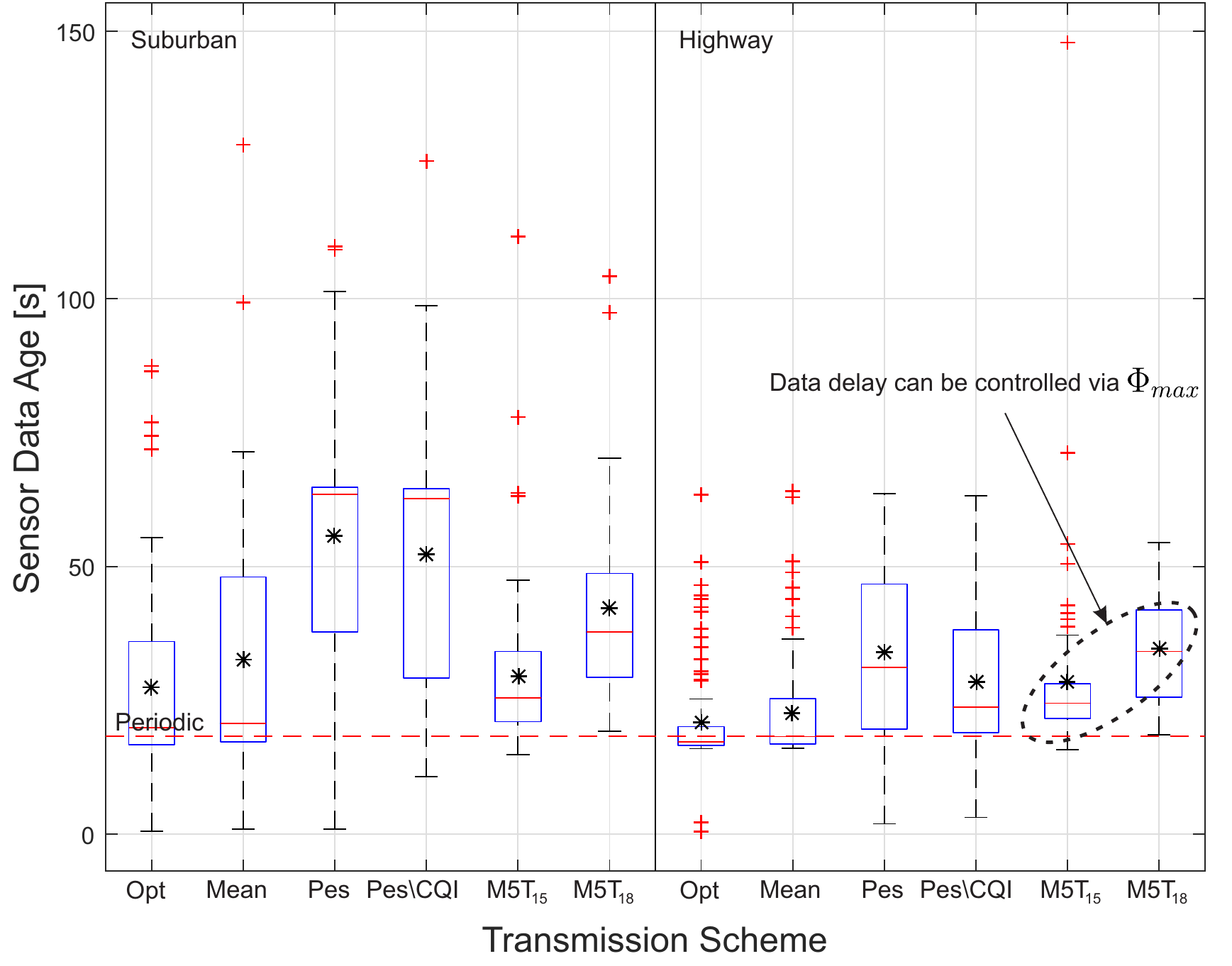}{0.45}{}
	
	\caption{End-to-end goodput and sensor data age for the considered multi-metric transmission schemes ($t_{min}=30$).}
	\label{fig:multi_metric}
	\vspace{-10pt}
\end{figure*}
Fig.~\ref{fig:multi_metric} shows the performance evaluation for the considered multi-metric transmission approaches. For this evaluation, the tracks are treated separately, as different behaviors were detected during the measurements. The weighted mean approach uses the same weights for all considered metrics as they achieved similar data rate values in the single-metric evaluations.
%
% Optimistic & mean: worse than single metric
%
It can be observed that the mean goodput for the \emph{optimistic} and \emph{weighted mean} approach is below the best single metric result which was achieved for \emph{$\Phi_{SNR}$}. It can be concluded that low values of a single metric can be considered meaningful indicators for unfavorable channel  connections. Contrastingly, high values of a single indicator do not provide enough information for detecting the presence of a connectivity hotspot.
%
% Pessimistic: CQI-dependent, t_max defines scheme behavior
%
For the \emph{pessimistic} approach, the results differ significantly depending on the considered evaluation track. While very high values are achieved for the highway track (mean: 14.58 MBit/s), the scheme shows a similar performance as the single-metric for the suburban track (mean: 11.16 MBit/s). The high rate of low \ac{CQI} values as shown in Fig~\ref{fig:correlation} combined with the low time variance due to relatively low velocity values results in a transmission scheme, which is mostly controlled by the transmission timeout $t_{max}$ and not the actual channel conditions.
%
% New pessimistic scheme without CQI
%
Therefore, another pessimistic transmission scheme is introduced, which does not integrate the \ac{CQI} information into the transmission decision. As shown in the figure, the resulting performance is similar for the two tracks (suburban mean: 12.63 MBit/s, highway mean: 12.65 MBit/s). 
%
% Conclusion: valleys vs hotspots, low vs high values
%
In conclusion, it is much more important to avoid transmissions during connectivity valleys than to leverage connectivity hotspots. 
%
% M5Tree
%
The best performance is achieved with the machine learning based approach which does not only achieve the highest mean data rate (18.5 MBit/s for the highway scenario with $\Phi_{max}$=18 MBit/s) but also reduces the number of occurrences of low data rate transmissions significantly. It is the only scheme that is not only considering the network quality, but also the size of the payload which has a decent influence on the achievable efficiency as shown in Fig.~\ref{fig:correlation}. The tradeoff between goodput and data delay can be controlled by the parameter $\Phi_{max}$.

\section{Conclusion}

%
% Introduction
%
In this paper, we presented a probabilistic model for the efficient transmission of sensor data in a vehicular context. 
%
% Problem statement
%
Its general goal is to find the optimal transmission time for vehicular sensor data with respect to the channel conditions in order to save communication resources and avoid retransmissions.
%
% Results: single-metric
%
In comprehensive field evaluations, the impact of different network quality indicators serving as metrics for the transmissions decision was evaluated. Although all considered metrics are able to increase the resulting data rate significantly, they differ in terms of variance and \ac{DMR}.
%
% Results: multi-metric
%
Through application of a machine learning based combination of the available network quality indicators, the average goodput can be further improved significantly while the delay is only slightly increased. Other variants of the proposed transmission scheme can be considered depending on the \ac{DMR} requirements of the application.
%
% Future work
%
In future work, we want to further improve the transmission scheme by taking interdependencies with the transport layer protocols into account. Furthermore, we want to enhance the accuracy of network quality estimation by integrating knowledge about the cell capacity obtained from passive control channel analysis as described in \cite{FalkenbergHeimannWietfeld2017}.

\section*{Acknowledgment}

%\footnotesize
Part of the work on this paper has been supported by Deutsche Forschungsgemeinschaft (DFG) within the Collaborative Research Center SFB 876 ``Providing Information by Resource-Constrained Analysis'', projects A4 and B4 and has been conducted within the AutoMat (Automotive Big Data Marketplace for  Innovative  Cross-sectorial  Vehicle  Data  Services)  project, which received funding from the European Union’s Horizon 2020 (H2020) research and innovation programme under the Grant Agreement No 644657. Thomas Liebig received funding from the European Union Horizon 2020 Programme (Horizon2020/2014-2020), under grant agreement number 688380  ``VaVeL: Variety, Veracity, VaLue: Handling the Multiplicity of Urban Sensors''.

\bibliographystyle{IEEEtran}
\bibliography{Bibliography}

\end{document}